\relax 
\providecommand{\mciteSetMaxWidth}[3]{\relax}
\providecommand{\mciteSetMaxCount}[3]{\relax}
\bibstyle{achemso}
\mciteSetMaxCount{main}{bibitem}{43}
\mciteSetMaxCount{main}{subitem}{1}
\mciteSetMaxWidth{main}{bibitem}{786432}
\mciteSetMaxWidth{main}{subitem}{0}

\documentclass[journal=mamobx,manuscript=article]{achemso}
\usepackage{graphicx,mciteplus}
\usepackage[version=3]{mhchem}

\title{End Grafted Polymer Nanoparticles in a Polymeric Matrix: Effect of Coverage and Curvature}
\author{Joshua Kalb}
\affiliation{Department of Chemical Engineering, Columbia University, New York, NY}
\author{Douglas Dukes}
\affiliation{Department of Materials Science \& Engineering, Rensselaer Polytechnic Institute, Troy, NY.}
\author{Sanat K. Kumar}
\email{sk2794@columbia.edu}
 \affiliation{Department of Chemical Engineering, Columbia University, New York, NY}
\author{Robert S. Hoy}
\affiliation{Departments of Mechanical Engineering and Physics, Yale University, New Haven, CT}
\author{Gary S. Grest}
\affiliation{Sandia National Laboratories, Albuquerque, NM}
\date{\today}
\begin{document} 

\begin{abstract}
It has recently been proposed that the miscibility of nanoparticles with a polymer matrix can be controlled by grafting polymer chains to the nanoparticle surface. As a first step to study this situation, we have used molecular dynamics simulations on a single nanoparticle of radius R ($4\sigma \le$R$\le 16\sigma$, where $\sigma$ is the diameter of a polymer monomer) grafted with chains of length 500 in a polymer melt of chains of length 1000. The grafting density $\Sigma$ was varied between $0.04$-$0.32$ chains/$\sigma^2$. 
To facilitate equilibration a Monte Carlo double-bridging algorithm is applied - new bonds are formed across a pair of chains, creating two new chains each substantially different from the original.
For the long brush chains studied here, the structure of the brush assumes its large particle limit even for $R$ as small as 8$\sigma$, which is consistent with recent experimental findings.
We study autophobic dewetting of the melt from the brush as a function of increasing $\Sigma$. Even these long brush and matrix chains of length $6$ and $12$ $N_e$, respectively,  (the entanglement length is $N_e \sim 85$) give somewhat ambiguous results for the interfacial width, showing that studies of two or more nanoparticles are necessary to properly understand these miscibility issues. Entanglement between the brush and melt chains were identified using the primitive path analysis.  We find that the number of entanglements between the brush and melt chains scale simply with the product of the local monomer densities of brush and melt chains.
\end{abstract}

\maketitle

\noindent\newline

\noindent{\bf 1. Introduction}

Controlling the interactions, and hence the spatial dispersion, of nanoparticles (NP) is critical to the ultimate goal of producing polymer nanocomposites with desired macroscale properties.\cite{mackay,akcora}
Experimental studies \cite{XXX,YYY,mackay,akcora} have shown that, most often, NPs tend to randomly aggregate into clusters or migrate to interfacial regions;  consequently, the advantages of their nanoscale dimensions are lost in terms of property improvements. Thus, the current barrier to integrating NPs into a range of advanced devices is the challenge of controllably dispersing and organizing them within durable matrices while retaining their unique properties.
One way to alleviate this issue is by end-grafting polymer chains to the NPs, a system which is commonly referred to as a polymer brush.\cite{akcora} For high enough coverage, the NPs are sterically stabilized and hence good nanoparticle spatial dispersion can result. \cite{3}

Previous work on planar polymer brushes in a chemically-identical matrix has shown that the matrix only wets the polymer brush if the melt chains are shorter than the brush.\cite{1,grest96,4,5,6,7} 
Longer melt chains are known to spontaneously dewet the brush (``autophobic dewetting''), because the entropic loss associated with the matrix chains penetrating the brush cannot overcome the translational, or mixing, entropy.\cite{3,8,harton} Thus, for a given graft density, increasing the chain length of the polymer matrix effectively changes the solvent quality experienced by the polymer brush from ``good'' to ``poor''.
Similar ideas \cite{Witt88,Wijm93,Zhul90,3} can be used to understand the miscibility of polymer grafted NPs in a polymer melt, except that, in addition to the graft density and chain lengths of the brush and melt, one must also account for the curvature of the NP. For small NPs (high curvature), the polymer brush chains can explore more space compared to a planar surface, resulting in less entropic loss to penetrate the brush, reducing the tendency for autophobic dewetting. 
Most previous computational studies of polymer brushes grafted to a nanoparticle have either not included the polymer matrix \cite{akcora,bindercurv} or have studied only short chains in which both brush and melt  are well below  the entanglement chain length, $N_e$.\cite{smith09} There have also been studies on bare spherical nanoparticles in long chain polymer matrices.\cite{depablo}


Perhaps the most commonly accessible measure of brush structure is its mean height, a quantity that has been the focus of several experimental studies. Chevigny et al. \cite{boue} recently used neutron scattering to measure the conformation of silica NPs with an average radius of 13 nm grafted with a polystyrene polymer brush dispersed in a polystyrene matrix. 
They found that if the melt chains become long enough, the brushes compressed by a factor of two in thickness compared to their stretched conformation in solution. Other experiments,\cite{ohno,savin,dukes} which also measure the mean brush height, have been able to critically comment on existing theories of curved polymer brushes. For example, theories that assume that all the free chain ends are a uniform distance from the grafting surface do not provide a good representation of the experimental results.\cite{Daou82,Alex77} Theories which relax this assumption\cite{Witt88,Wijm93,Zhul90} 
provide a much better description of the experimental data. These theories also predict that there is a exclusion zone for the free ends - that is, the ends cannot approach the particle surface, thus creating an ``exclusion zone''. This result is in sharp contrast to planar brushes, where no exclusion zone is predicted from theory or found in experiments.

While previous work has focused on a single moment of the spatial distribution of monomers, the mean brush height (the first moment), examining the full distribution of monomers is a more sensitive and critical test of the theories. This type of comparison is best performed via simulation studies, which we undertake here. To provide context, we consider the theoretical predictions of radial distance dependence of the density of grafted chain monomers.   

In the poor-solvent limit, Daoud and Cotton predicted that the brush chains collapse into a globule whose chain density profile is simply a constant value of $\frac{|v|}{2\omega}$, where $\nu$ and $\omega$ are the second and third virial coefficients respectively.\cite{Daou82}  This result is independent of the curvature of the particles and reflects the dominance of the solvent conditions alone.
For $\Theta$ and good solvents, in the limit of large curvature (small particle radius, $R$), the chain density profile is predicted to scale as \cite{Wijm93}
\begin{equation}
	\phi(r)\propto [\Sigma^{1/2}(R/r)]^m
	\label{eq:largecurvprofile}
\end{equation}
where $r$ is the radial distance from the center of the particle, $\Sigma$ is the graft density and the exponent, $m$, is 1 or 4/3 for a $\Theta$ or good solvent, respectively.  Note that $\phi$ is nonlinear in $\Sigma$ because the brush height $H$ depends on $\Sigma$.  
Implicit in this treatment is the unrealistic assumption that all chain ends are extended to an identical degree, and fixed to a ``phantom surface''; an assumption that is not representative of physical systems. In contrast, in the limit of small curvature (large $R$) one must account for a distribution of chain ends, e.g., as proposed by Milner-Witten-Cates.\cite{Miln88} Following this form, the segment density profile is parabolic, given by\cite{Wijm93}
\begin{equation}
	\phi(z)=\frac{3\Sigma Na^3}{H_o}\left(\frac{H}{H_o}\right)^2\left(1-\left(\frac{z}{H}\right)^2\right)
	\label{eq:smallcurvprofile}
\end{equation}
\noindent where $a$ is the statistical segment length, $z=r-R$ is the radial distance from the particle surface, $H_o$ is the effective brush height for a flat surface, and $H$ is the brush thickness.  Note that this form is reminiscent of planar polymer brushes and suggests that, asymptotically, we must recover the planar brush result for arbitrarily large particles.

Finally, for intermediate particle sizes, a combination of large and small curvature behaviors is conjectured.  Thus, the segment density profile has large curvature behavior near the surface [\ref{eq:largecurvprofile}], followed by small curvature behavior away from the surface.  The crossover between the two behavior occurs at $z_0$, which is obtained by matching the two solutions. The combined profile is \cite{Wijm93}
\begin{equation}
\phi(z)=\left\{
\begin{array}{l l}
  {\left(\frac{3}{64\pi^2}\right)^{1/3}\frac{f^{2/3}\nu^{-1/3}a^{4/3}}{(R+z)^{4/3}}} & \quad \mbox{$0\leq z\leq z_o$}\\
  \frac{3\pi^2}{16a^2N_2^2\nu}(H_2^2-(z-z_o)^2) & \quad \mbox{$z_o\leq z \leq H_2+z_o$}\\ \end{array} \right. 
  \label{transitionprofile}
\end{equation}
where $f$ [$\equiv \Sigma (4\pi R^2)$] is the number of grafted chains, $\nu$ is the second virial coefficient, $z_o$ is the location of the transition, $N_2$ is the number of segments beyond $z_o$, and $H_2$ is the brush thickness beyond $z_o$. 

In addition to the form of the monomer distribution function, these theories also predict the density profiles of the (free) chain ends. The Daoud-Cotton model assumes that all the chain ends are a uniform distance away, implying a delta function distribution. The Wijmans-Zhulina model, the spherical brush analog of the Milner-Witten-Cates model, has a well defined exclusion zone (where the probability of finding a chain end becomes negative and hence unphysical). The expression for the density of free ends is given by eq. 19 in ref. \cite{Wijm93}.

For the long chains modeled in this study, we can readily determine if entanglements occur between the brush chains and the matrix chains. Such information is useful in determining the role of these entanglements in the mechanical reinforcement gained by a grafted particle immersed in a polymer melt. A particularly useful concept in this context is that of the primitive path \cite{Doi86}, which gives a measure of how the presence of other (long) chains forces the central chain to be transiently confined in a ``tube".  
Using the primitive path analysis (PPA) developed by Everaers and co-workers,\cite{Ever04} it is now possible to get an approximate measure of an individual chain's primitive path, entanglements, and hence interactions with all other chains in the melt via relaxation methods of individual melt configurations.

In this work, we use molecular dynamics (MD) simulations to model spherical nanoparticles for three values of radii, $R$, and four coverages, $\Sigma$, coated with end-tethered chains of length $N_t$, immersed in a melt of chains of length $N_m$, where both $N_t$ and $N_m$ are much greater than the entanglement length $N_e$.
In the next section, we present the polymer model and simulation methodology. We then present results for the monomer density profiles for both the brush and melt chains. 
The conformations of the brush and melt chain are examined at various scales and compared to theory.
Finally, using PPA analysis,\cite{Ever04,Hoy07} we examine the properties of the system entanglements within the polymer brush and between the polymer brushes and the melt.  

\vspace{0.2in}
\noindent{\bf 2. Model and Simulation Methodology}

\noindent
{\bf 2.1 Interaction potentials and system parameters.}
End grafted polymer coated NPs in a polymeric matrix are simulated using the coarse grained bead spring model of Kremer and Grest.\cite{Krem90} All brush and melt monomers have a mass $m$ and interact via the truncated and shifted Lennard-Jones potential
\begin{equation}
U_{LJ} = \Bigg\{
\begin{array}{lcc}
  {4\epsilon\left( \left(\frac{\sigma}{r_{ }}\right)^{12}-\left(\frac{\sigma}{r_c}\right)^{12}-\left(\frac{\sigma}{r_{ }}\right)^{6}+\left(\frac{\sigma}{r_c}\right)^{6}         \right)} & , & \quad \mbox{$r<r_c$}\\
  0 & , & \quad \mbox{$r>r_c$}\\ \end{array}
\label{eq:ULJ}
\end{equation}
where $\sigma$ it the monomer diameter and $\epsilon$ is the characteristic pair interaction energy.
Covalently bonded monomers also interact via the finitely extensible nonlinear elastic (FENE) potential
\begin{equation}
U_{FENE} = -\displaystyle\frac{kR_0^{2}}{2} ln\left(1 - (r/R_0)^2\right),
\label{eq:ufene}
\end{equation}
where $k = 30k_BT$ and $R_o=1.5\sigma$ \cite{grest96}. 
The NP is modeled as a spherical particle of radius $R$ consisting of a uniform distribution of monomers of diameter $\sigma$ and density $\rho_{NP}$.
The interaction between the NP and the brush and melt monomers is given by integrating \ref{eq:ULJ} over this sphere \cite{intveld2009}: 
\begin{equation}
\begin{array}{lcl}
U_{\rm NP,p}(r) & = & \frac{2 ~ R^3 ~ \sigma^3 ~ A_{\rm NP,p}}{9 \left( R^2 - r^2 \right)^3} \bigg[1 -\\
& & \frac{\left(5 ~ R^6+45~R^4~r^2+63~R^2~r^4+15~r^6\right) \sigma^6}
                                {15 \left(R-r\right)^6 \left( R+r \right)^6}\bigg]        
\end{array}
\label{eqn5}
\end{equation}
where $A_{\rm NP,p} = 24\pi \epsilon_{\rm np} \rho_{NP} \sigma^3$.
Here $\epsilon_{\rm np} = \epsilon$ is the interaction between a polymer monomer and a monomer in the nanoparticle. 
$\rho_{\rm NP}=1.0\sigma^{-3}$ gives $A_{\rm NP,n} = 75.4 \epsilon$. The NP-monomer interaction is truncated at $r_c=(R+\sigma)$.

For all systems, $n_t$ chains were randomly tethered at a coverage $\Sigma$ to the surface of the NP, with the only restriction that no two grafting points lie within $2\sigma$ of each other. 
Three values of the NP radius were studied, $R =4$, $8$, and $16\sigma$, for coverages 
$\Sigma= 0.04$, $0.08$, $0.16$ and $0.32$ chains/$\sigma^2$. The number of tethered chains varied from $8$ to $1024$ depending on the coverage and NP radius. The number of melt chains $n_m=2000$ for $R=4\sigma$, $3000$ for $R=8\sigma$ and $5000$ for $R=16\sigma$. 
Each melt chain has a degree of polymerization $N_m= 1000$, while each tethered chain  
is of length $N_t=501$.
For flexible bead spring chains, the entanglement length $N_e$ for bulk melts is $N_e =85\pm 7$ \cite{hoy09}, so the polymer melt is well entangled.

For NP-brush-melt systems, the initial states were created as described in Auhl {\it et al.}\cite{auhl03} with the additional restriction that no monomers overlap with the NP which is held fixed at the center of a periodic, cubic $L\times L \times L$ simulation cell. $L$ was initially chosen so that overall monomer density $\rho_{\rm m}= 0.75\sigma^{-3}$. 
Overlapping monomers in the initial states were pushed off each other using a soft potential until the monomers were far enough apart to switch on the full Lennard-Jones interaction. After the full Lennard-Jones potential was turned on the system was equilibrated at a pressure $P=5.0\epsilon/\sigma^3$ which for this system is the pressure of a neat polymer melt of long chains at a monomer density $\rho_{\rm m}= 0.85\sigma^{-3}$.\cite{Krem90}  
The equilibrated size of the simulation cell is $L \sim 132\sigma$ for $R=4\sigma$, $\sim 154$ for $R=8\sigma$ and $\sim 182\sigma$ for $R=16\sigma$, with slight variations depending on the coverage $\Sigma$.
The number of melt chains was sufficiently large to ensure that the tethered chains do not interact with their periodic images.

The systems in this study were thoroughly equilibrated using the double-bridging method described in Refs.\cite{karayiannis02,auhl03,sides04,Hoy07}.
Because diffusive equilibration of entangled brush-melt systems is exponentially slow, 
Monte Carlo moves that alter the topological connectivity of chain subsections were periodically performed, allowing the chain configurations to relax more rapidly.\cite{karayiannis02,auhl03}

We also simulated NP-brushes in an implicit solvent, as has been done in previous simulations of a dry brush on a flat surface,\cite{mura89,gres93} a cylindrical surface\cite{mura91} and a sphere.\cite{bindercurv} 
All simulations of the NP-brush-melt systems and NP good-solvent brushes are run at $T = \epsilon/k_B $ and $r_c= 2^{1/6}\sigma$.
We also simulated NP-brushes in an implicit poor solvent with $T=1.5\epsilon/k_B$ and $r_c=2.5\sigma$.
For this model, the $\Theta$ temperature is $T_\Theta\sim 3.18\epsilon/ k_B$.\cite{graessley99}

\begin{figure}[htbp]
\includegraphics[width=5in]{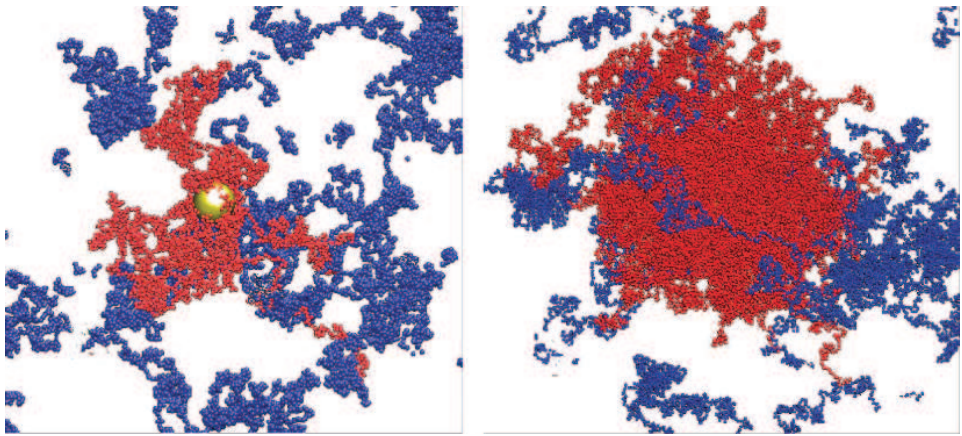}
\caption{Snapshot of a NP of radius $R=4\sigma$ for (a) $\Sigma=0.04$ and (b) $\Sigma=0.32$ chains/$\sigma^2$. Most of the melt chains (blue) have been removed to visualize the grafted chains (red). }
\label{fig:nanoimage}
\end{figure}

All simulations were carried out using the LAMMPS parallel molecular dynamics package. \cite{plimpton95} 
Newton's equations of motion were integrated with a velocity-Verlet algorithm with a time step $\delta t=0.012 \tau$ for the melt and good solvent simulations and $0.005\tau$ for the poor solvent, where $\tau=\sigma(m/\epsilon)^{1/2}$. All monomers were coupled to a Langevin thermostat with a damping constant $\Gamma=0.5 \tau^{-1}$. \ref{fig:nanoimage} shows representative snapshots from the simulations, where it is clear that there is intimate contact between the brush chains and the surrounding melt. Increasing the graft density serves to sterically shield the NP surface, suggesting that brush physics should ultimately decide the thermodynamic state (miscible vs. clustered particles) of these composites.

\noindent
{\bf 2.2 Primitive Path Analysis}
For NP-brush-melt systems, we performed a primitive path analysis (PPA) to identify the entanglements between the brush chains and between the brush and melt chains.\cite{Ever04,Hoy07}.
All chain ends are fixed in space, and several changes are made to the interaction potential. 
Intrachain excluded-volume interactions are set to zero, while interchain excluded-volume interactions are retained. The covalent bonds are strengthened by setting $k=100\epsilon/\sigma^2$. 
FENE bonds are monitored to ensure they do not exceed 
$1.2\sigma$; this prevents chains from crossing.\cite{Suku05}
Self-entanglements are not necessarily preserved, but their contribution is typically negligible\cite{Suku05,Silb82,Silb88}. 
The system is coupled to a heat bath at $T=0.001\sigma/k_B$,
and the equations of motion are integrated until the chain lengths are minimized. 

\begin{figure}[htbp]
\includegraphics[width=3.0in,angle=270]{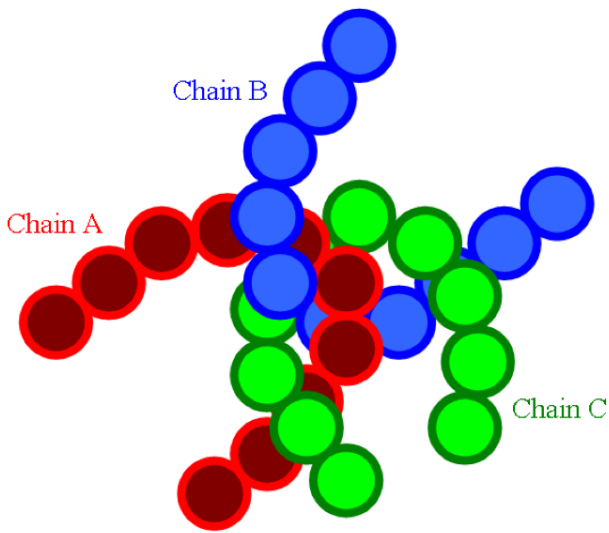}
\caption{Schematic of entanglement identification. For visualization purposes, chain A is red, chain B is  blue, and chain C is green.}
\label{fig:entident}
\end{figure}

Since the standard PPA is not sufficient to identify individual entanglements, following the convergence of the standard PPA, we therefore perform ``thin-chain PPA'' using the same method described in Ref.~\cite{Hoy07}.
At the end of the standard PPA, $n_{dec} - 1$ new beads are inserted between adjacent beads on the original chains. Here, $n_{dec}=8$. The bead diameter (i.e., range of the LJ potential), range of Lennard Jones interaction between the nanoparticle and system monomers, and $R_0$ are reduced to $1/n_{dec}$ of their initial values. Covalent bonds are further strengthened by setting $k=100 n_{dec}/\sigma^2$, and their lengths are capped at $1.2\sigma/n_{dec}$ to avoid chain crossings.   The equations of motion are integrated until the chains again minimize their length. Results presented here are averaged over 10 to 20 statistically independent systems.

 \ref{fig:entident} schematically depicts our PPA analysis procedure. 
One entanglement between chains A and B is identified as a block of consecutive monomers on chain A having interchain contacts with chain B. 
As described in Ref.~\cite{Hoy07} identifying these blocks is somewhat subtle; we identify interchain ``contacts'' as monomers whose smallest interparticle neighbor distance is less than $r_{con}=(5/4)(1/n_{dec})2^{1/6}\sigma$. For computational simplicity, we follow convention and assume all entanglements are binary, i.e., involve only two chains. Ternary entanglements (a vertex including three chains A, B, C) are counted as three entanglements: $A-B$, $A-C$, and $B-C$ [\ref{fig:entident}]; this may overestimate the absolute entanglement densities, but the relative values of brush-brush versus brush-melt entanglement densities should be accurately captured.

\vspace{0.2in}
\noindent{\bf 3. Chain Structure}

\begin{figure*}
\centering
\includegraphics[width=6.5in]{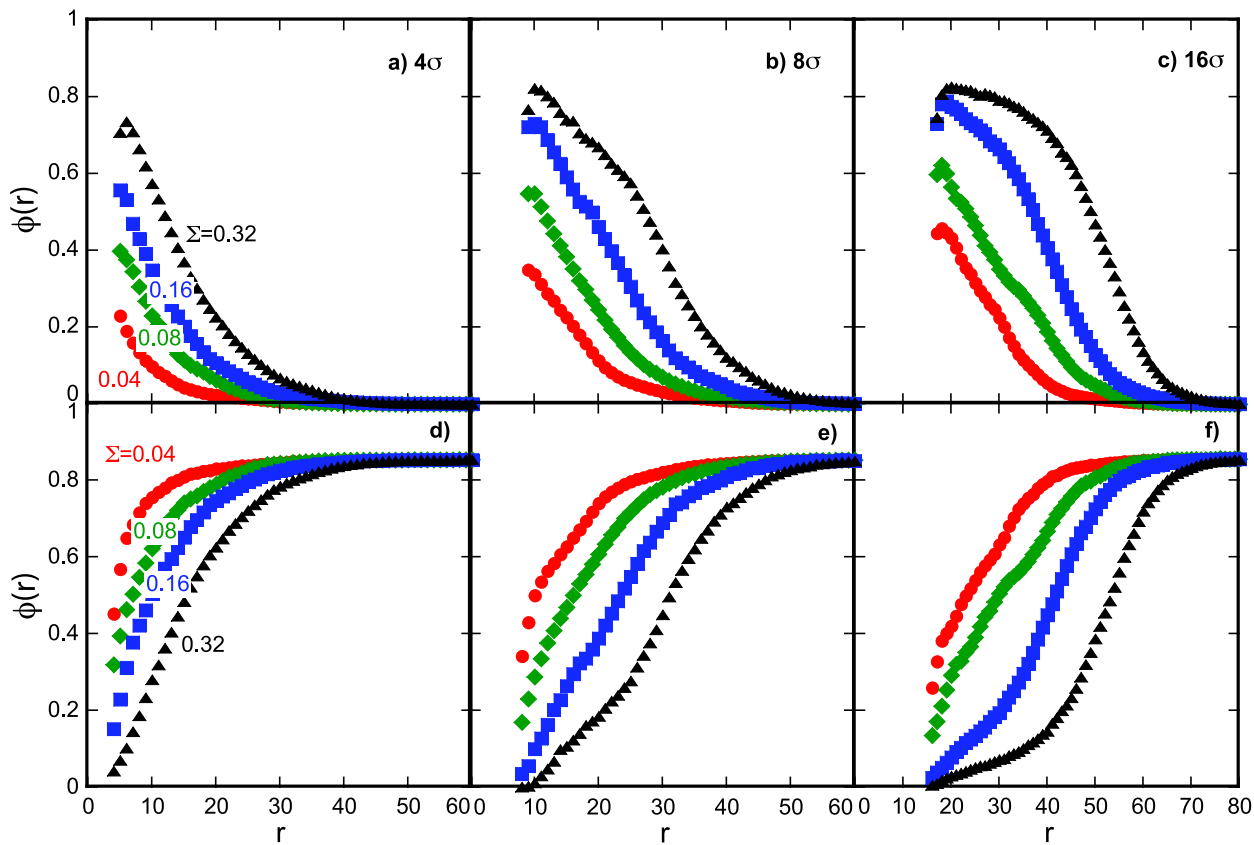}
\caption{Monomer density profiles for NP radius $R=4\sigma$ (parts a and d), $8\sigma$ (b and e) and $16\sigma$ (c and f). The density of grafted chains is shown in the top row while the density of melt chains is shown in the bottom row.  }
\label{fig:monomerdensity}
\end{figure*}

The radial density, $\phi(r)$, of the tethered chains (\ref{fig:monomerdensity} a-c) and the corresponding melts (\ref{fig:monomerdensity} d-f) are shown for three different nanoparticle radii $R=4$, $8$, and $16\sigma$, respectively, and four different grafting densities,  $\Sigma= 0.04$, $0.08$, $0.16$, $0.32$ chains/$\sigma^{2}$.  
It is apparent that the radial monomer density profile of the brushes extends farther into the melt with higher grafting densities. Further, the shape of the monomer density profile changes quite dramatically as the NP radius increases from $R=4$ to $R=16\sigma$. The smallest NPs show a concentration profile that is more concave, while for the largest NP the profiles are more parabolic in its shape, especially for intermediate values of $r$. 

The melt density profile (\ref{fig:monomerdensity}d-f) typically mirrors the density properties of the grafted brushes. The sum of the two densities suggests that the system is essentially incompressible, except in the immediate vicinity of the grafting surface.  This result is not surprising given the high overall density of the system. 

\begin{figure*}
\includegraphics[width=6.5in]{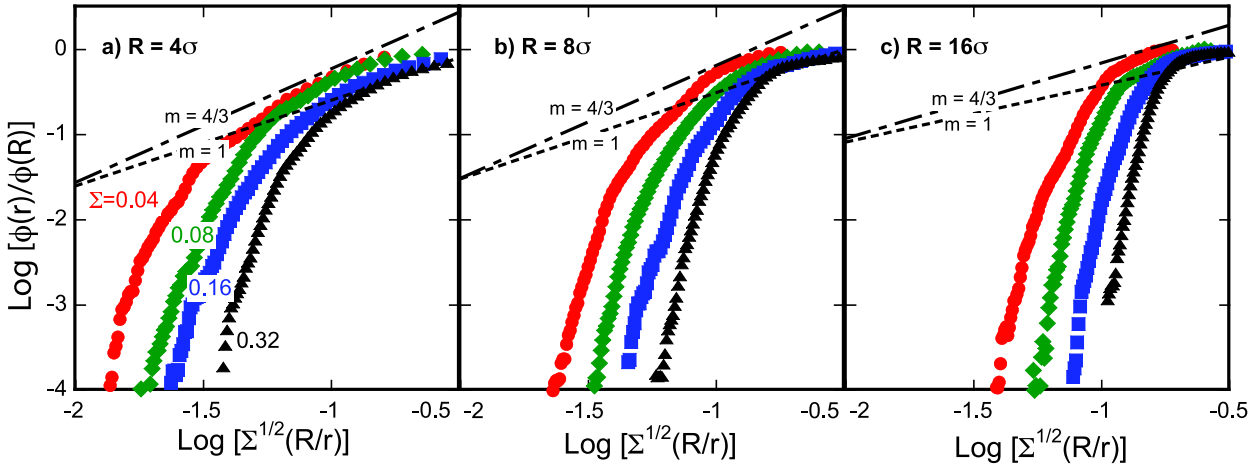}
\caption{Comparison of the normalized segment density profiles to large curvature behavior for a) $R=4\sigma$, b) $R=8\sigma$, and c) $R= 16\sigma$.  Slopes of 1 and 4/3 (cf. \ \ref{eq:largecurvprofile}) are drawn as a guide to the eye.}
\label{fig:largecurvprofile}
\end{figure*}  

To understand the dependence of the monomer density profile on the $\Sigma$ and $R$, we perform an analysis inspired by Daoud-Cotton, i.e., following  \ref{eq:largecurvprofile}. \ref{fig:largecurvprofile} shows that this scaling only approximately describes the density profile at small $r$ or large $R/r$. For all particle diameters and small $\Sigma$ it appears that a scaling exponent of $m=4/3$ approximately describes the data near the particle surface.  As $\Sigma$ increases, the solvent quality appears to get progressively poorer, evidenced by the decay in the scaling exponent.  This suggests a transition from ``good'' to ``poor'' solvent conditions with increasing grafting density. Note, however, that these ideas are only qualitative since there is no region where the scaling relationship is followed exactly; rather, the data are always curved. 
For large $r$ (small $R/r$), however, the data do appear to be better fit by an approximately exponential behavior reminiscent of that found for flat surfaces - this behavior presumably follows from the finite chain lengths, and hence finite extensibility, of the graft chains simulated.  We can thus conclude that the model of Daoud and Cotton only qualitatively describes the simulation results for the brush density profile. However, it evidently shows that solvent quality becomes worse as one considers more densely grafted particles, a situation in which one may expect a smaller propensity for the melt chains to wet the brush.

\begin{figure*}
\includegraphics[width=6.5in]{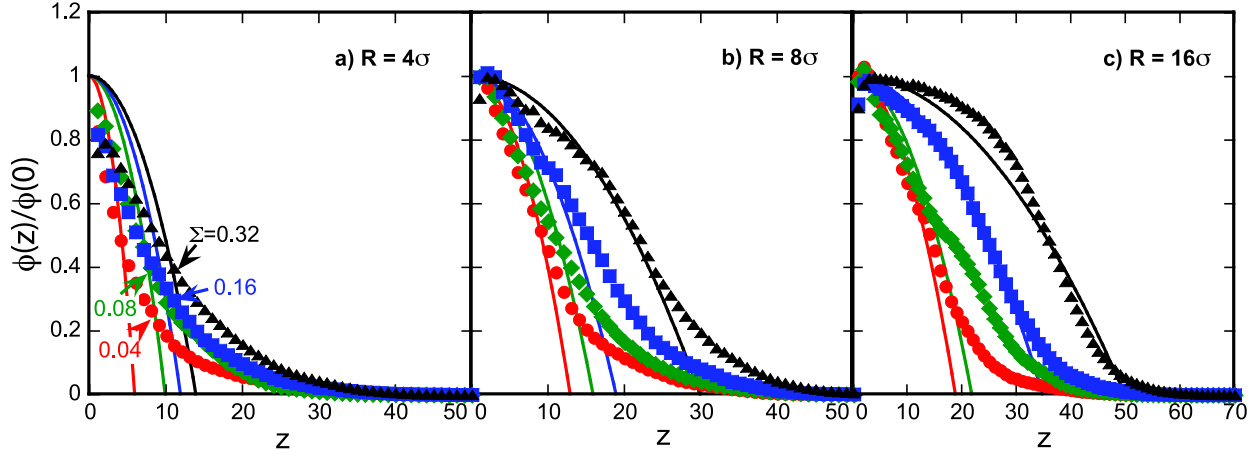}
\caption{Comparison of the normalized segment density profiles to small curvature behavior (\ \ref{eq:smallcurvprofile}) for a) $R=4\sigma$, B) $R=8\sigma$, and C) $R=16\sigma$.} 
\label{fig:parabolascaling}
\end{figure*}

\begin{figure*}
\includegraphics[width=6.5in]{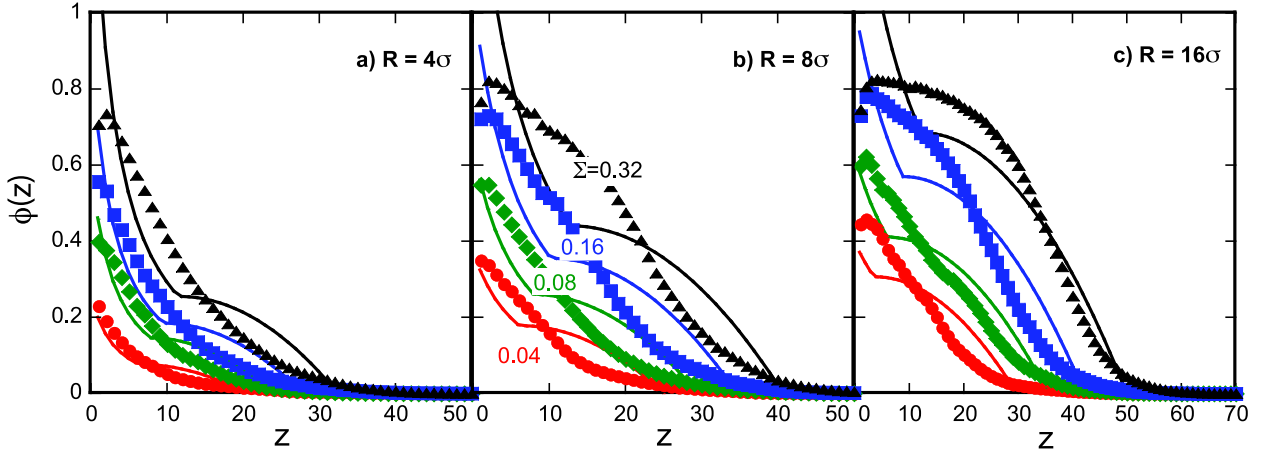}
\caption{Comparison of the normalized segment density profiles to intermediate curvature behavior (\ref{transitionprofile}) for a) $R = 4\sigma$, b) $R=8\sigma$, and c) $R=16\sigma$.}
\label{fig:transitionalprofile}
\end{figure*}

To obtain a better understanding of the brush density profile, we also examined the applicability of the large particle (large $R$) ideas embodied in \ref{eq:smallcurvprofile}. \ref{fig:parabolascaling} shows that \ref{eq:smallcurvprofile} surprisingly describes the results obtained for all $R$ except for $R=4\sigma$.  While this treatment describes the behavior for all small $r$ values for $R\ge 8\sigma$, apparently for all $\Sigma$ values we consider, poor agreement is found for larger $r$ values. As described above, this discrepancy probably results from the finite extensibility of the chains, a factor not built into these theories. Consistent with these findings \ref{fig:transitionalprofile} shows that the transitional form \ref{transitionprofile}, which is supposed to apply for intermediate $R$ values, does not provide improved fits.  
An important conclusion we draw from this analysis is that the large particle limit is attained relatively quickly, and particles as small as 10 times the monomer diameter of the chains are already in this limit.

\begin{figure}
\includegraphics[height=6.5in]{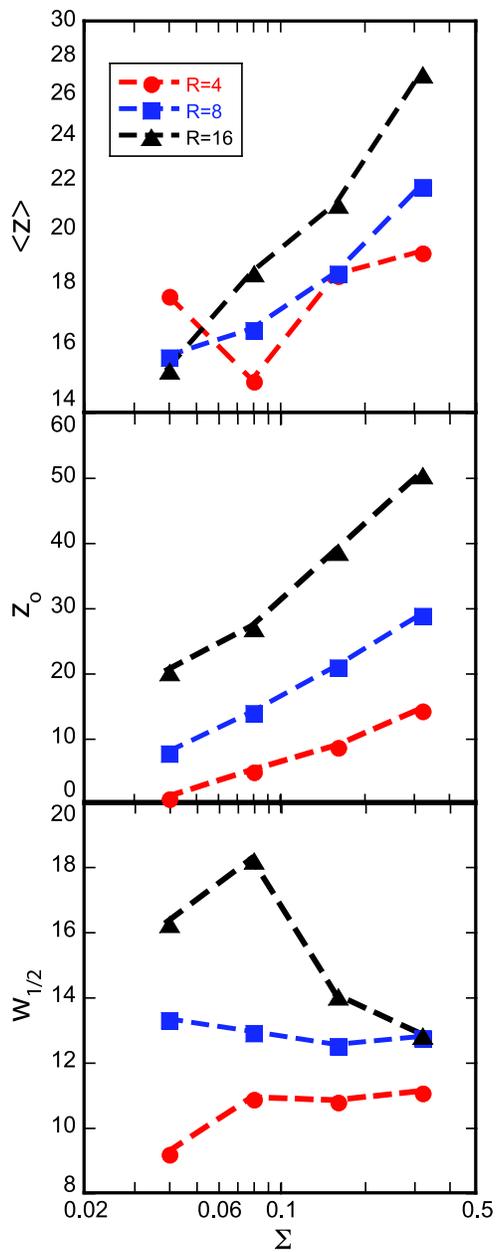}
\caption{(a) Mean brush height $<z>$, (b) interface position $z_o$, and (c) interface width $w_{1/2}$ as a function of coverage $\Sigma$ for NPs of radius $R=4$ (circles), $8$ (squares), and $16\sigma$ (triangles).}
\label{fig:fighscale}
\label{fig:scale}
\end{figure}

A few experimental points are considered in the context of these simulation results.  One question: is the apparent good-solvent behavior found for the largest $R$ reasonable, given that the brush should ``collapse" due to the autophobic dewetting that is expected for long enough matrices and high enough grafting densities?  Note here that all of the monomer density data for $R\ge 8\sigma$ appear to be fit by a parabolic profile which is indicative of good solvent behavior. Thus, at first glance there appears to be no signature of the dewetting phenomena in these monomer density profiles. We then conjectured that, perhaps, a moment of the monomer density, namely the mean height $<z>$ of the polymer brush,
\begin{equation}
\left<z\right>=\frac{\int (z+R)^2 z \phi(z) dz}{\int (z+R)^2 \phi(z) dz}
\label{eq:height}
\end{equation}
\noindent might give better insights into this dewetting phenomenon (\ref{fig:fighscale}). 
In addition to this ``direct" measure of height, another measure can be derived by fitting the monomer density profile to well known functional forms. In particular, we follow an approach that we previously used to analyze the effective $\chi$ parameter between the brush chains and the matrix.\cite{harton} We first define a normalized density,
$\rho=\phi_{graft}(z)/(\phi_{graft}(z)+\phi_{matrix}(z))$.  
In ref.~\cite{harton} 
we showed that concentration profiles derived from self-consistent mean-field theory could be fit to a standard form: $\rho=\frac{1}{2} \left[1- tanh \left(\frac{z-z_o}{w_{1/2}}\right)\right]$, where $z_o$ is the position of the interface and $w_{1/2}$ is the interfacial width. Paralleling the approach for phase segregated diblocks, then, $w_{1/2}=(6\chi)^{-1/2}$. This form fits all of our data very well, and in \ref{fig:scale} we report both the $z_o$ and $w_{1/2}$ values derived from the fits. Clearly $z_o$ tracks the behavior seen for the mean brush height $<z>$. The results (\ref{fig:fighscale}a) show that for the largest radius particle, $<z>\sim \Sigma^{1/2}$, is consistent with the expectation of a stretched brush in a good solvent. In contrast, the $R=4\sigma$ data are very noisy, but are essentially independent of $\Sigma$. More interestingly, the interface width $w_{1/2}$, in all cases, converges to a value of around 10-12$\sigma$ at the largest surface coverages, $\Sigma$, examined. 
This number is comparable to the radius of gyration of the melt chains, $R_g \approx \sqrt{\sigma l_K N_{m}/6} \approx 17\sigma$ (where the Kuhn length $l_K \simeq 1.8\sigma$), suggesting that we are limited by the size of the chains in question. 
An unambiguous measurement of interfacial width, without interference from this length scale, requires that the two length scales be very different. Even though the chains employed here are relatively long, they are still too short to allow for this separation of length scales and hence we cannot determine if the brush and melt chains are phase-separating due to autophobic dewetting.

\begin{figure*}
\includegraphics[width=3in]{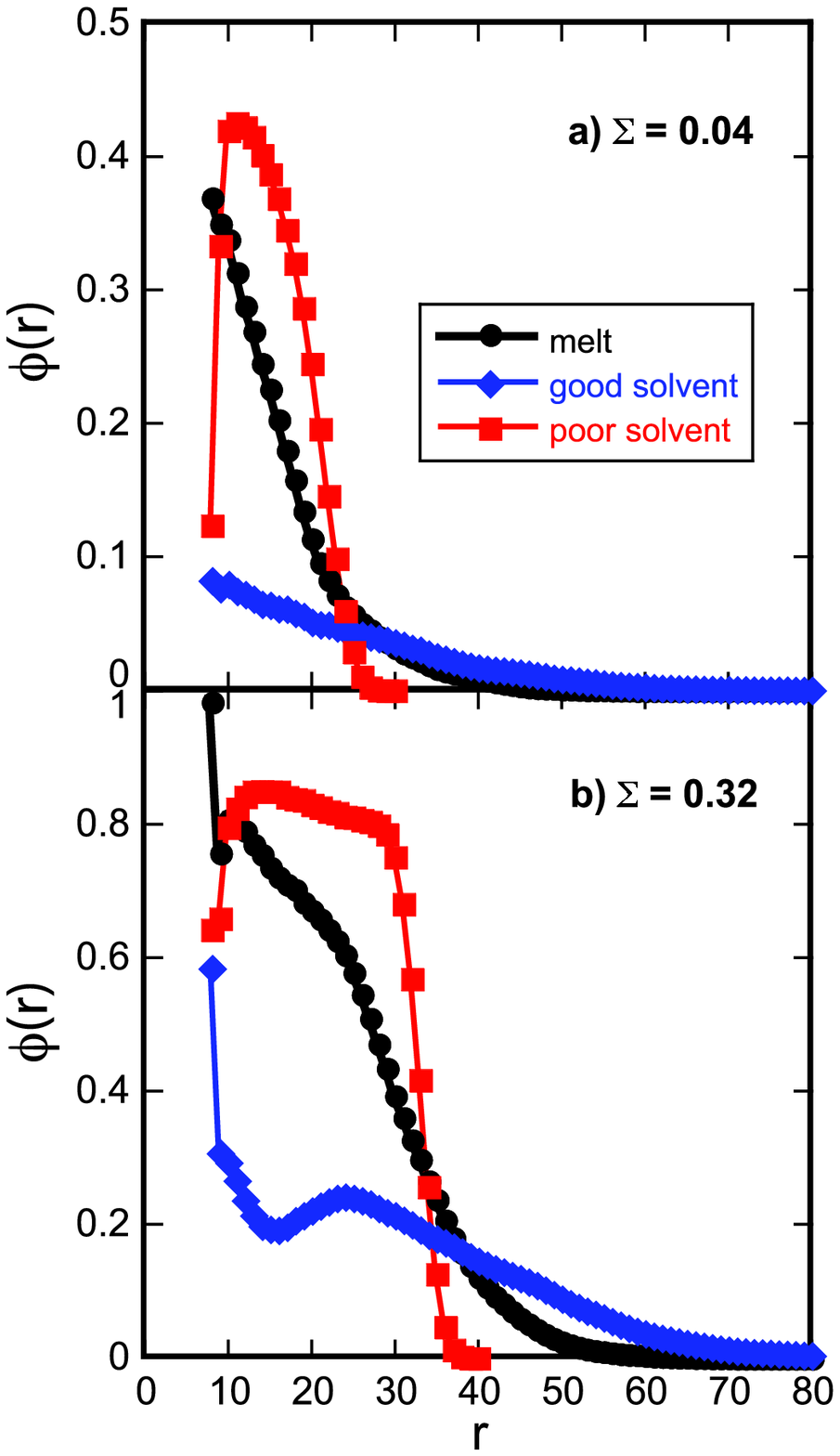}
\caption{Brush monomer density for NP of radius $R=8 \sigma$ for (a) $\Sigma=0.04$ chains/$\sigma^2$ and (b) $0.32$ chains/$\sigma^2$ in a polymer melt and in an implicit good and poor solvent.}
\label{fig:solventquality}
\end{figure*}

In this context, it is also informative to compare the melt systems with simulation in an implicit good and poor solvent.  For a typical system (see \ref{fig:solventquality}), close to the surface, we see that the curved brushes interacting with a melt have a density profile that resembles more that seen for an implicit, poor solvent than for a good solvent. However, there is a much longer tail for the brushes immersed in a melt. Similar behavior was also previously observed for brushes grafted to a plane;
in both cases the `long tail' is probably attributable to finite chain length effects.

\begin{figure*}
\includegraphics[width=6.5in]{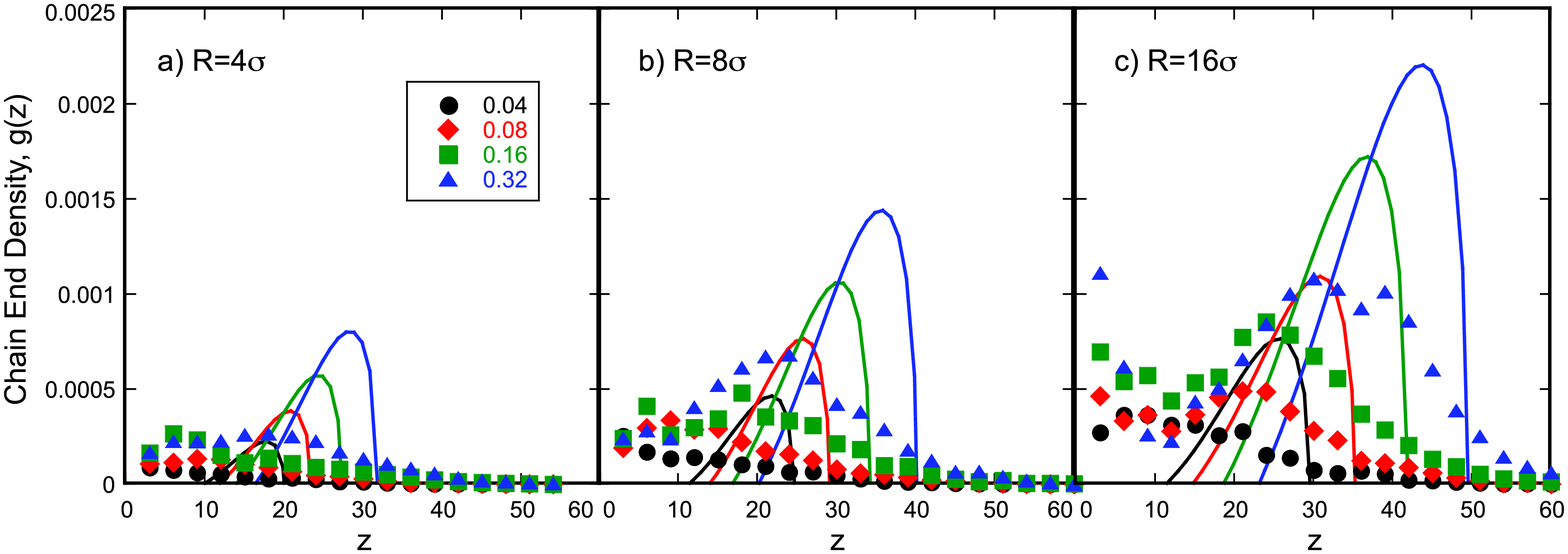}
\vspace{0in}
\caption{End density $\rho_e(z)$ of the tethered chains in the presence of melt chains for NP radius (a) $R=4\sigma$, (b) $8\sigma$, and (c) $16\sigma$ for four values of the coverage $\Sigma$. The lines are predictions of the Wijmans-Zhulina theory.\cite{Wijm93}}
\label{fig:enddensity}
\end{figure*}

A particularly interesting prediction of mean-field theories of curved brushes is the existence of a zone of exclusion for the free chain ends near the surface. To study this interesting possibility we examine the free chain end distributions, $\rho_e(z)$ of the polymer brush in \ref{fig:enddensity}. It is apparent that there is a maximum in this quantity, but that there is no complete exclusion of chain ends from near the particle surfaces. These results can be compared to the predictions of the Wijmans-Zhulina theory in the limit of large particles. The theory does however captures the maximum in the end density profile at some distance away from the particle surface and its relative position with increasing graft density.

\begin{figure*}
\centering
\includegraphics[width=6.5in]{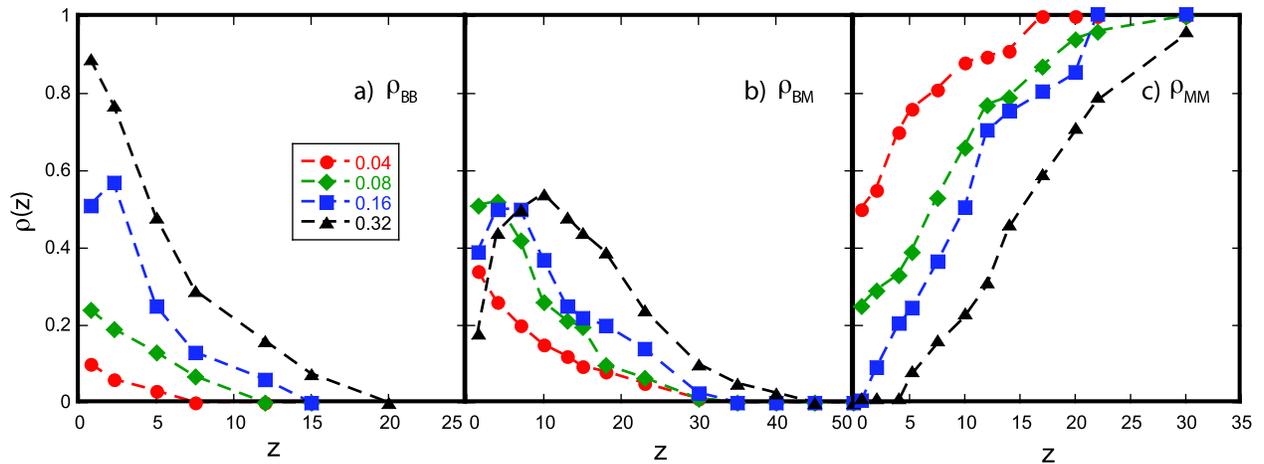}
\vspace{0.0in}
\caption{Scaled densities for (a) brush-brush entanglements, (b) brush-melt entanglements, and (c) melt-melt entanglements for a NP of radius $R=4\sigma$. All distances are measured from the NP surface.  Results are averaged over 10-20 independent initial configurations.}
\label{fig:entanglementsscaled4}
\end{figure*}

Finally we identified the entanglements between the polymer brush and the melt.
\ref{fig:entanglementsscaled4} shows results for the brush-brush, brush-matrix and matrix-matrix entanglement densities, $\rho_{BB}(z)$, $\rho_{BM}(z)$ and $\rho_{MM}(z)$ respectively scaled by the melt density $\rho_{\rm m}$ for $\Sigma=0.04-0.32$ chains/$\sigma^2$ and $R=4\sigma$. 
The densities $\rho_{BB}(r)$ and $\rho_{BM}(r)$ increase
with increasing coverage $\Sigma$. The maximum entanglement density, which occurs at the same $r$ where the total monomer density also has a peak, scales as $\Sigma^{1.4}$.
As previously observed for planar brushes,\cite{Hoy07} the fraction of brush-melt entanglements scales to a good approximation as the product of the monomer densities, $\phi_{brush}(r)\phi_{melt}(r)$, as expected for 'binary' entanglements.
\ref{fig:entanglementsscaled4}b shows that the fraction of brush-melt entanglements peaks at a distance $3-10\sigma$ away from the surface of the nanoparticles. The $z$ position of the maximum increases with increasing $\Sigma$.
Similarly, the fraction of melt-melt entanglements (see \ref{fig:entanglementsscaled4}c) also scales roughly quadratically with the melt monomer density, and increases monotonically with increasing $z$.
Given the simple behavior observed, we did not consider the larger radii where similar results are expected.
Results for for planar brushes are reported
in Ref.~\cite{Hoy07}.

\vspace{0.2in}
\noindent{\bf 4.0 Conclusions}

We have performed detailed MD simulations of the behavior of a polymer brush grafted to a single NP in contact with a long chain melt. Our results have a number of salient points. (i) We show that the brush monomer density for NPs of radius $R=4\sigma$ is close to the behavior anticipated for a star polymer, while, for radii $R=8$ and $16\sigma$ the brush monomer density is already close to that of the large particle limit. This result is consistent with experimental results. (ii) The conformational transition between a stretched brush and a collapsed brush, which has been attributed to the autophobic dewetting of the brush by the melt, is not readily apparent in the brush density profiles. We conjecture that this difficulty is due to the finite size of the chains in question, even though the number of beads in the simulated brush chains is long enough to be well entangled -  we, thus, cannot cleanly resolve the interfacial width attributed to dewetting effects for the native sizes ($R_g$) of the chains in question. Simulations of two or more grafted nanoparticles, which will help to unequivocally resolve this issue, are currently underway. (iii) The brush end-monomer density profiles are qualitatively consistent with the large nanoparticle  
predictions of Zhulina and Wijmans, though we do not see any clear evidence for a zone of depletion. 
This may also be a manifestation of the finite chain lengths considered in our work. (iv) The spatially dependent entanglement densities semiquantitatively scale as a product of the respective densities of the two chains that form the brush-brush, brush-melt and melt-melt entanglements respectively. 

\vspace{0.2in}
\noindent{\bf Acknowledgements}

We thank the New Mexico Computing Application Center NMCAC for a generous allocation of computer time. Funding for this research at Columbia was provided by the National Science Foundation through a NIRT grant. Support from NSF Award Nos.\ DMR05-20415, DMR-0835742 and an Anderson Fellowship from Yale University is gratefully acknowledged.  
This work was performed, in part, at the Center for Integrated Nanotechnologies, a U.S. Department of Energy, Office of Basic Energy Sciences user facility. Supported, in part, by the Laboratory Directed Research and Development program at Sandia National Laboratories.  Sandia is a multiprogram laboratory operated by Sandia Corporation, a Lockheed Martin Company, for the United States Department of Energy under Contract No. DEAC04-94AL85000. This material is based upon work supported, in part, by the New York State Office of Science, Technology \& Academic Research under contract no. C070119.


\end{document}